\documentclass[11pt]{article}

\usepackage{appendix}
\usepackage{amssymb,amsmath,amsfonts,dsfont}
\usepackage{graphicx}
\usepackage{subfig}
\usepackage[figuresright]{rotating}
\usepackage[applemac]{inputenc}
\usepackage{graphicx}
\usepackage{dcolumn}
\usepackage{bm}
\usepackage{amscd,amsthm}
\usepackage{ifthen}
\usepackage{booktabs}
\usepackage{bm}
\usepackage{makecell} 

\usepackage{mdframed}
\usepackage{listings} 
\usepackage{xcolor} 
\usepackage{tcolorbox} 

\usepackage[skip=0pt]{caption}

\usepackage{hyperref}
\hypersetup{
	bookmarks=true,         
	unicode=false,          
	pdftoolbar=true,        
	pdfmenubar=true,        
	pdffitwindow=false,     
	pdfstartview={FitH},    
	pdfauthor={Tobit Klug},     
	pdfsubject={Subject},   
	pdfcreator={Tobit Klug},   
	pdfproducer={Producer}, 
	pdfnewwindow=true,      
	colorlinks=true,       
	linkcolor=red,          
	citecolor=green,        
	filecolor=magenta,      
	urlcolor=cyan           
}
\usepackage{breakurl}

\usepackage[
backend=biber,
url=false,isbn=false,doi=false,eprint=false,
date=year,
hyperref=auto,
style=alphabetic,
natbib=true,
sorting=nty,
giveninits=true,
maxnames=2, maxbibnames=10,
]{biblatex}

\AtEveryBibitem{%
  \clearname{translator}%
  \clearlist{publisher}%
  \clearfield{pagetotal}%
  \clearname{editor}
  \clearfield{pages}
  \clearfield{series}
  \clearlist{address}
  \clearfield{address}
  \clearname{address}
  \clearname{volume}
  \clearfield{volume}
  \clearname{number}
  \clearfield{number}  
} 

\usepackage{tikz}
\usepackage{mathtools}
\usepackage{pgfplots}
\usepgfplotslibrary{groupplots} 
\usetikzlibrary{pgfplots.groupplots}
\usetikzlibrary{matrix,arrows,decorations.pathmorphing}
\usepackage{pgfplots,pgfplotstable}
\usepgfplotslibrary{fillbetween}
\usetikzlibrary{shadows,arrows}

\usepackage{changepage} 

\usetikzlibrary {arrows.meta} 
\usetikzlibrary{fit} 
\usetikzlibrary{backgrounds}

\usepgfplotslibrary{colorbrewer}
\pgfplotsset{compat=newest}
\usetikzlibrary{external}

\usetikzlibrary{calc}
\usetikzlibrary{positioning}

\usepackage{graphics}
\usepackage{comment}
\usepackage{readarray} 
\usepackage{caption} 
\usepackage{xcolor}         
\usepackage{footnote}
\makesavenoteenv{tabular}
\usepackage{refcount}

\usepackage{hyperref}       
\usepackage{url}            
\usepackage{amsfonts}       
\usepackage{nicefrac}       
\usepackage{microtype}      

\usepackage{afterpage}

\newdimen\nodeDist
\usepackage{url}
\usepackage{breakurl}
\usepackage{outlines} 
\usepackage{changepage} 

\usepackage{rays_defs_18}

\definecolor{darkgreen}{RGB}{50, 168, 82}

\newcommand{\dimz}{k_z}
\newcommand{\numFk}{k_x}
\newcommand{\numPk}{k_y}
\newcommand{\numQk}{k_z}
\newcommand{\numFd}{r_z}
\newcommand{\numPd}{r_x}
\newcommand{\numQd}{r_y}

\begin{document}

\begin{center}
	
	{\bf{\LARGE{
                Reliable Evaluation of MRI Motion Correction: Dataset and Insights
	}}}
	
	\vspace*{.2in}
	
	{\large{
			\begin{tabular}{cccc}
			Kun~Wang$^{\ast}$, Tobit~Klug$^{\ast}$, Stefan~Ruschke$^\dagger$, Jan S.~Kirschke$^\dagger$, and Reinhard~Heckel$^{\ast,\star}$
			\end{tabular}
	}}
	
	\vspace*{.05in}
	
	\begin{tabular}{c}

 $^\ast$School of Computation, Information and Technology, Technical University of Munich \\
 $^\dagger$School of Medicine and Health, Technical University of Munich \\
 $^\star$Munich Center for Machine Learning
	\end{tabular}
	
	\vspace*{.1in}
 
	\today
	
	\vspace*{.1in}
	
\end{center}

\begin{abstract}
Correcting motion artifacts in MRI is important, as they can hinder accurate diagnosis. 
However, evaluating deep learning-based and classical motion correction methods remains fundamentally difficult due to the lack of accessible ground-truth target data. 
To address this challenge, we study three evaluation approaches: real-world evaluation based on reference scans, simulated motion, and reference-free evaluation, each with its merits and shortcomings. 
To enable evaluation with real-world motion artifacts, we release PMoC3D, a dataset consisting of unprocessed \textbf{P}aired \textbf{Mo}tion-\textbf{C}orrupted \textbf{3D} brain MRI data. 
To advance evaluation quality, we introduce MoMRISim, a  feature-space metric trained for evaluating motion reconstructions. 
We assess each evaluation approach and find real-world evaluation together with MoMRISim, while not perfect, to be most reliable. 
Evaluation based on simulated motion systematically exaggerates algorithm performance, and reference-free evaluation overrates oversmoothed deep learning outputs.
\end{abstract}

\section{Introduction}

\label{sec:intro}

Magnetic resonance imaging (MRI) is a powerful medical imaging modality. However, its long scan times make it  sensitive to motion artifacts caused by patient movement.
Motion artifacts can prevent effective diagnosis and may force repeating the scan~\cite{andreQuantifyingPrevalenceSeverity2015,slipsagerQuantifyingFinancialSavings2020}.  
Both classical and increasingly deep learning-based algorithmic approaches have been proposed for correcting artifacts resulting from rigid motion in 2D~\cite{haskellNetworkAcceleratedMotion2019,singhJointFrequencyImage2022,singhDataConsistentDeep2023,levacAcceleratedMotionCorrection2023,wuMonerMotionCorrection2024} and 3D~\cite{corderoSensitivityEncodingAligned2016,johnsonConditionalGenerativeAdversarial2019,duffyRetrospectiveMotionArtifact2021,almasniStackedUNetsSelfassisted2022,klugMotionTTT2DTestTimeTraining2024} MRI.

However, research on 3D motion correction is challenging as the field lacks a standardized evaluation approach for realistically evaluating different approaches. The core issue is that ground-truth target data is fundamentally difficult or impossible to obtain:
\begin{itemize}

\item Real-world motion-corrupted data captures true motion but lacks ground-truth for quantitative assessment. To enable real-world evaluation, one can collect two scans, a motion-corrupted and a motion-free one, and use the motion-free as a target or ground-truth. However, the two scans need to be aligned, require careful pre-processing, and the motion-free scan might also be slightly motion corrupted.  

\item Most commonly, evaluation is conducted based on simulated rigid motion artifacts in which case computing reference-based metrics is straight forward. 
However, evaluation might be unrealistic due to the motion simulation. Potential artifacts from non-rigid motion due to brain pulsation and from second-order motion effects such as motion-induced magnetic field inhomogenities or spin history effects are not accounted for when simulating motion~\cite{spiekerDeepLearningRetrospective2024}.
In addition, data has to be fully sampled to simulate motion, which is rarely the case for 3D MRI.  

\item Finally, evaluation can be conducted with reference-free image quality metrics, which avoid the need for pre-processing or the lack of a realistic reference. However, classical gradient-based metrics  correlate poorly with perceived image quality~\cite{marchettoAgreementImageQuality2024}. 

\end{itemize}

In this work, we advance evaluation by systematically assessing real-world, simulated, and reference-free evaluation, and by providing PMoC3D, a real-world dataset for evaluation of 3D motion correction methods.

Our dataset, PMoC3D, consists of raw \textbf{P}aired \textbf{Mo}tion-\textbf{C}orrupted \textbf{3D} brain MRI data for enabling real-word evaluation by comparing reconstructions motion-corrupted data to a reference scan.  
For PMoC3D, we collected three motion-corrupted scans of different motion severity each from eight subjects along with one motion-free scan to use as a reference. 

First, we study real-world evaluation with a reference scan based on the PMoC3D dataset by assessing how well reference-based image quality metrics correlate with human assessment. We consider standard metrics in the pixel-space such as SSIM~\cite{wangSSIM2004} and PSNR~\cite{horeImageQualityMetrics2010} and feature-space metrics such as 
DreamSim~\cite{fu2023dreamsim}, and we additionally  propose a feature-space metric MoMRISim that is trained to align with varying levels of motion severity. 
We find that, for scans with moderate to severe motion corruption, reference-based evaluation using feature-space metrics like MoMRISim correlates well with human judgments and provides a reliable measure of reconstruction quality.

However, under mild motion, the motion-free reference reconstructions often retain residual artifacts, and in some cases,  mildly motion-corrupted scans reconstructed with motion-correction methods appear visually cleaner than the reference. 
This challenges the reliability of reference-based evaluation in mild motion settings, where simulated data with known ground truth can offers a more meaningful alternative for evaluation.

Second, we assess evaluation based on simulated motion  corruption, and observe that some methods achieve almost error free reconstructions under the most severe simulated motion, 
whereas the same methods exhibit noticeable residual artifacts under severe real-world motion.
This is consistent with findings for other imaging problems, that found simulated data to potentially lead to missleading conclusions~\cite{shimron_implicit_2022}.

Third, regarding reference-free evaluation, we propose and consider a vision-language model (VLM) score. While exhibiting a significantly better alignment with perceived image quality than classical gradient-based reference-free metrics, we find the VLM score to be biased towards reconstructions, which are overly smooth but potentially miss anatomical details. 

In summary, all three considered evaluation methods have shortcomings, but evaluation on real-world paired datasets such as PMoC3D, when combined with an appropriate feature-based metric such as MoMRISim, provides a relatively reliable and meaningful assessment of reconstruction performance under moderate to severe motion. 

\section{The PMoC3D Dataset for real-world evaluation}
\label{sec:pmoc3d_dataset}

In order to evaluate accelerated 3D motion correction methods, we constructed the PMoC3D dataset, described in this section. PMoC3D is a 3D dataset containing the raw measurement data of scans with real-world motion as well as corresponding motion-free scans as a target or estimate of the ground-truth. 

Previous works relied on evaluation datasets that provide only the processed magnitude images~\cite{johnsonConditionalGenerativeAdversarial2019,duffyRetrospectiveMotionArtifact2021,ganzDatasetsDeliberateHead2022,liMoCoDiffAdaptiveConditional2024}.  
However, approaches that explicitly estimate motion in order to correct for it, require access to the unprocessed raw data (k-space data) \cite{corderoSensitivityEncodingAligned2016,klugMotionTTT2DTestTimeTraining2024}.

\subsection{Acquiring paired motion-free and motion-corrupted data}

For the PMoC3D dataset we scanned 8 healthy male and female subjects. 
The study was approved by the institutional review board and written informed consent was obtained from all participants prior to data collection (see the final paragraph of this subsection for the full ethical statement).
From each subject we acquired four scans, one motion-free and three motion-corrupted labeled as S\{subject\}\_\{scan\} with subjects $1,\ldots,8$ and scans $0,\ldots,3$, where 0 corresponds to the motion-free case. 
When we observed artifacts in the scanner's reconstruction while recording a motion-free scan, we assumed that the subject has moved and restarted the scan session. 
We provide access to the corrupted scans (S3\_4, S5\_4, S8\_4) with involuntary motion resulting in a total of 27 motion-corrupted scans.

\paragraph{Data acquisition.} The data was acquired on an Ingenia Elition 3.0T X scanner (Philips Healthcare, Best, The Netherlands) with the standard 16-channel dStream HeadSpine coil array (of which 13 channels were automatically selected based on SNR). 
We performed Cartesian 3D T1-weighted Fast-Gradient-echo (TFE) imaging with a 1mm isotropic resolution and a field of view of $221 \times 170 \times 256$ mm. 
With $\numQk$ and $\numFk$ oversampling of $1.4$ and $2.0$, the acquisition matrix-size is $\numPk \times \numQk \times \numFk = 222 \times 236 \times 512$. 

Data was acquired with undersampling along the phase encoding dimensions $\numPk \times \numQk$ (axial plane) with an undersampling factor of $\mu=4.94$, a densely sampled auto-calibration region of size $37\times37$ and partial-Fourier sampling with factor 0.85 in the $\numQk$ direction(see Figure~\ref{fig:data_display} in Appendix~\ref{appsec:dataset_details} for the resulting undersampling mask).

We provide access to the full-size k-space data.
To reduce the computational cost for  evaluating experiments, 
we also 
cropped the data, where we crop along the fully-sampled read-out dimension $\numFk$ to the size of the field of view (256) by subsampling every second voxel.
The sequence parameters are in Appendix~\ref{appsec:dataset_details}.

\begin{figure}
    \centering
    \includegraphics[width=0.9\linewidth]{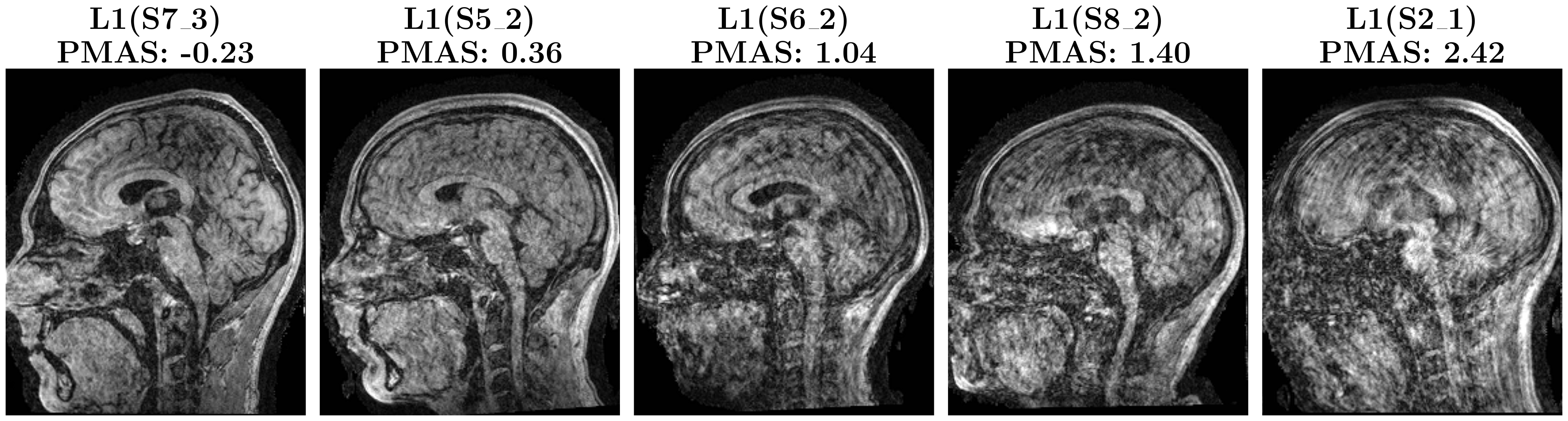}
    \caption{Sagittal views and corresponding perceived motion artifact scores (PMAS) of selected L1-reconstructions from our dataset with varying degrees of motion artifacts ranging from mild (\textbf{S7\_1}, \textbf{S5\_1}) to moderate 
    \textbf{S6\_2}) to severe (\textbf{S8\_2}, \textbf{S2\_1}). These examples highlight the challenges in reconstructing motion-corrupted scans.
} 
    \label{fig:L1_show_severity}
\end{figure}

\paragraph{Sampling trajectory.} The Cartesian k-space is acquired within $N_s=52$ shots resulting in $\numPk\ast\numQk/(\mu N_s)=222\ast236/(4.94 * 52) = 204$ acquired read-out lines per shot. The acquisition of one shot lasts 1.35s followed by a pause of 1.74s resulting in a total scan duration of 2:40min. The read-out lines are acquired following a quasi-random sampling trajectory except the $3\times3$ center of the k-space which is acquired at the start of the first shot. Hence, the sampling order varies between all scans in the dataset. 

We chose a random order because it ensures that both low-/high-frequency components are sampled in every shot which is beneficial for motion estimation~\cite{corderoSensitivityEncodingAligned2016,usmanRetrospectiveMotionCorrection2020,klugMotionTTT2DTestTimeTraining2024}.

\paragraph{Sensitivity maps.}  For each subject, the dataset contains the coil sensitivity maps calculated using Gyrotools MRecon~\cite{gyrotools_reconframe} through a calibration scan performed at the beginning of each subject's scan session for which the subject was instructed to hold still. 

\paragraph{Motion.} Motion corrupted scans are obtained by instructing the subject to perform a motion at one or more time instances during the scan. The instructions are as follows: 

\begin{itemize}
    \item Slightly turn head left/right and stay/return to origin.
    \item Nodding: Once up, once down and return to origin.
    \item Head shaking: Once left, once right and return to origin. 
    \item Move chin towards chest and stay/return to origin. 
\end{itemize}
We instruct to perform a motion slowly if it should be performed slowly as opposed to abruptly. 

To generate a diverse dataset containing mild to severe motion artifacts we vary the instructions itself, when we give them, and how many we give (up to three). 
The instructions and time stamps are provided with the data.

\paragraph{Ethical statement.}The local institutional review board approved the study (study number 2024-365-S-NP) in accordance with the ethical standards of the institutional and/or national research committee and with the 1964 Helsinki Declaration and its later amendments or comparable ethical standards. Prior informed consent was obtained from all individual participants.
 
\subsection{Categorizing motion artifacts in the data}
\label{sec:pmoc_motion_severity}

Stronger motion and more motion events make reconstruction more difficult. To facilitate evaluation with the PMoC3D dataset, we provide a quantitative measure of human-perceived severity of the motion artifacts of each of the 24 scans corrupted with voluntary motion. 

The score is computed as follows. 
We first obtain the L1-reconstruction with wavelet regularization~\cite{lustigCompressedSensingMRI2008} without any motion correction for each scan (see Appendix~\ref{appsec:L1_details} for details).
The L1-reconstruction with wavelet regularization does not account for motion modeling or correction, allowing us to directly evaluate the severity of motion artifacts in each scan.
The first two authors of this paper perform pairwise comparisons between the reconstructions classifying either one to have more or both to have similarly severe artifacts. 
Based on these pairwise results we fitted a Bradley--Terry model~\cite{bradley1952rank} to obtain a perceived motion artifact score for each scan indicating the perceived severity of motion artifacts relative to all other scans in the dataset (see Appendix~\ref{sec:hps_cal} for more details). 

Figure~\ref{fig:L1_show_severity} shows sagittal views of L1-reconstructions from our dataset from left to right in ascending order according to their perceived motion artifact score. 
As we can see, the severity of motion artifacts increases with increasing perceived motion artifact score. 
Reconstructions S7\_3 and S5\_2 show mild artifacts, where most brain anatomical details are preserved despite the presence of minor ringing artifacts. 
In reconstruction S6\_2, the artifacts are more pronounced and obscure finer details, while in S8\_2 and S2\_1 the artifacts are severe enough that the brain structures become barely discernible. 
These examples illustrate the range of challenges encountered when reconstructing motion-corrupted scans in our dataset.

% ...

\section{Evaluation approaches}
\label{sec:eval_approaches}
This section presents three evaluation approaches 
for evaluating 3D MRI motion correction methods: real-world evaluation with a reference scan, evaluation based on simulated motion, and reference-free evaluation. 
We also discuss standard evaluation metrics and two novel metrics we propose for the respective approaches. 

\subsection{Real-world evaluation with a reference scan}

We perform real-world evaluation with our PMoC3D dataset, which consists of paired acquisitions for each subject. Each subject Si (i=1,$\ldots$,8) undergoes one motion-free scan Si\_0 and three motion-corrupted scans Si\_j (j=1,2,3). 
The motion-corrupted scans are categorized into two difficulty levels: the 8 scans with the lowest perceived motion artifact scores (PMAS) are labeled as mild motion-corrupted scans, while the remaining scans are classified as moderate and severe.
Each baseline method is applied to each motion-corrupted scan, and the L1 reconstruction of the motion-free scan is the reference for quantitative scoring. 

We preprocess the data to mitigate the challenges of comparing two different acquisitions, as suggested by the paper~\cite{marchettoAgreementImageQuality2024}.  
First, using advanced normalization tools(ANTs)~\cite{tustisonANTsXEcosystemQuantitative2021} rigidly aligns the motion-corrupted volume to the reference. 
Subsequently, a brain mask is generated via BET~\cite{smithFastRobustAutomated2002} on the motion-free scan and applied to both datasets, to focus the evaluation on the anatomical region of interest. 
Both volumes are multiplied by the brain mask and normalized to a max value of the 99.9th percentile before computing scores. 

After preprocessing, reference-based quality metrics are computed between the corrected motion-corrupted scans and the L1-based motion-free reference reconstructions (Si\_0).

\subsection{Evaluation based on simulated motion} 
\label{subsec:eval_sim_desc}
We evaluate performance by generating synthetic motion-corrupted, undersampled k-space data from the fully sampled Calgary Campinas Brain MRI dataset~\cite{souzaOpenMultivendorMultifieldstrength2018}, and comparing reconstructions against the original, uncorrupted reference volumes. 
This has the advantage that we have accurate ground-truth or target information, and the disadvantage that the motion is synthetic. 

A 3-D Cartesian mask with an acceleration factor of 4.9 is applied in the two phase-encoding directions, replicating the mask geometry used for PMoC3D. Each acquisition is divided into 52 shots following a random trajectory, again mirroring the paired real-world protocol. Inter-shot head motion is generated with an event-based framework designed to resemble PMoC3D artifacts. 
Motion events follow the instructions in Section~\ref{sec:pmoc3d_dataset} (head turning, nodding, etc.) which involve rotations about the $\numPk$-$\numQk$ and $\numFk$-$\numQk$ axes.
To more realistically capture head motion, we introduce random perturbations to the remaining motion parameters, i.e., the three translational components and the third rotational axis. These perturbations account for natural subject-specific variability and for the fact that real-world head rotations often occur around off-center axes rather than the image origin, resulting in complex motion patterns.
Motion severity is controlled by the number of events and their amplitude:  
\begin{itemize}
    \item \textbf{Mild}: One event with primary motion sampled uniformly from $\pm5^{\circ}$ and perturbations up to $\pm1^\circ$/$\mathrm{mm}$.
    \item \textbf{Severe}: Three events with primary motion sampled uniformly from $\pm15^{\circ}$ and perturbations up to $\pm5^\circ$/$\mathrm{mm}$.
\end{itemize}

For each baseline, we evaluate performance across ten test volumes. For every volume, we simulate the two severity levels, and for each level, we draw two independent motion seeds. Reconstructed volumes are normalized to their 99.9th percentile intensity, after which reference-based metrics are computed against the fully sampled ground-truth images.
While simulated motion offers controllable and repeatable conditions, it may oversimplify the motion dynamics in vivo.

\subsection{Reference-free evaluation}

In the reference-free setting, each baseline method reconstructs the motion-corrupted scans Si\_j, for i in 1,...,8, and j from 1 to 3 on PMoC3D.  
The reconstructed volumes are first masked to exclude non-brain regions, consistent with the paired evaluation protocol. Subsequently, the volumes are then normalized by the 99.9th percentile normalization, matching the normalization used in the paired evaluation.  
Because the available motion-free scans cannot serve as true ground truth, the normalised reconstructions are evaluated directly with a reference-free metric discussed in the next section.

\subsection{Evaluation metrics} 
\label{subsec:eval_metric}

To assess the effectiveness of motion correction methods with a reference, we used the reference-based metrics discussed below (that require access to a motion-free target image), and to assess without access to a reference, we use the reference-free metrics discussed below.

\paragraph{Reference-based metrics.} We consider standard pixel-wise metrics including Structural Similarity Index (SSIM)~\cite{wangSSIM2004}, Peak Signal-to-Noise Ratio (PSNR)~\cite{horeImageQualityMetrics2010}, and Artifact Power (AP)~\cite{jiPULSARMatlabToolbox2007}. We also consider perceptual metrics, which evaluate image quality based on high-level visual representations rather than direct pixel-wise comparisons. These metrics leverage deep learning-based feature extraction to assess structural and perceptual similarity, 
and have demonstrated strong correlation with radiologists' assessments~\cite{adamsonUsingDeepFeature2023}. Specifically, we consider 
Deep Image Structure and Texture Similarity (DISTS)~\cite{ding2020iqa} and DreamSim~\cite{fu2023dreamsim}.

Existing image quality metrics show varying degrees of effectiveness in evaluating motion artifacts in 3D MRI. Pixel-wise metrics such as PSNR and SSIM exhibit only moderate correlation with human evaluation. Perceptual metrics, while more aligned with human visual assessment, are generally trained on natural images and are not specifically designed for evaluating motion artifacts in MRI.

We propose \textbf{MoMRISim}, a similarity metric designed to quantify motion artifacts in 3D MRI. Building on the triplet-based training approach employed in DreamSim~\cite{fu2023dreamsim}, MoMRISim is trained using triplets consisting of one reference image and two motion-corrupted reconstructions, with a binary label indicating which reconstruction is perceptually closer to the reference. 
Rather than relying on manual annotations, we leverage simulated motion severity levels to generate supervision automatically, allowing us to scale training without human annotators. 
Triplets are constructed by applying synthetic rigid motion of varying severity, defined following the protocol in paper~\cite{klugMotionTTT2DTestTimeTraining2024}, to the fully sampled Calgary Campinas Brain MRI dataset~\cite{souzaOpenMultivendorMultifieldstrength2018}. 
To promote robustness across reconstruction styles, we randomly apply either L1-based or U-Net-based reconstruction methods without motion correction. This encourages the model to learn motion-artifact features that are invariant to the two reconstruction pipelines used in this study. Full training details, including data preparation and hyperparameters, are provided in Appendix~\ref{app:MoMRISim}.

\paragraph{Reference-free metrics.} We also investigate reference-free metrics for assessing image quality without requiring access to a motion-free reference scan.
We employ two gradient-based methods, Average Edge Strength (AES)~\cite{pannetier2016quantitative} and Tenengrad (TG)~\cite{kecskemeti2018robust}, which quantify image sharpness and structural clarity.

Classical gradient-based image quality metrics have been shown to correlate poorly with human judgments in prior work~\cite{marchettoAgreementImageQuality2024}. Our own experimental results in Appendix~\ref{appsec:corr_analysis} corroborate this finding, highlighting the limitations of these metrics in evaluating motion artifacts in 3D MRI.
To address this, we propose a novel reference-free \textbf{VLM score} based on prompting a vision-language model, specifically GPT-4o~\cite{openai_gpt4o,openai2024gpt4ocard}. The model receives reconstructions from three views (axial, coronal, and sagittal) and is asked to assign a motion artifact severity score ranging from 0 (no motion) to 3 (severe motion). To enhance robustness, we prompt the model independently five times at a temperature of 0.5 and compute the final score as the average across all runs.  
See Appendix~\ref{sec:gpt_evaluation} for details.

\section{Assessing evaluation approaches}
\label{sec:experiments}

In this section, we assess the three evaluation approaches. 





\subsection{Implementation details and baselines} 
\label{subsec:baselines}

We consider the following three motion reconstruction methods to generate reconstructions for evaluation:
\begin{itemize}
\item The classical
\textit{alternating optimization} \cite{corderoSensitivityEncodingAligned2016} 
alternates between two steps of L1-minimization reconstruction with wavelet regularization while fixing the motion parameters and four steps of motion parameter estimation while fixing the current reconstruction. 

\item The deep learning-based \textit{MotionTTT} \cite{klugMotionTTT2DTestTimeTraining2024} relies on a 2D U-net~\cite{ronnebergerUNet2015} pre-trained to perform motion-free MRI reconstruction. 
MotionTTT estimates the 3D motion parameters by optimizing a data-consistency (DC) loss between the network output and the given motion-corrupted measurements over the motion parameters (i.e., performing test-time-training (TTT)). 
The rational of the method is that because the network was trained on motion-free data, the DC loss is small when the motion parameters are estimated correctly. 

\item \textit{E2E Stacked U-nets} \cite{almasniStackedUNetsSelfassisted2022} is based on training a network consisting of a stack of refinement U-nets to predict the motion-corrected reconstruction from the motion-corrupted input end-to-end (E2E) with a single forward pass. E2E Stacked U-nets processes magnitude image volumes slice-wise to leverage the speed of 2D image-to-image networks, and use neighbouring slices as context to account for the 3D nature of the problem. 
During training we simulate motion as described in the Appendix~\ref{appsec:E2E_details} to generate training pairs of motion-free target and motion-corrupted input images. 
\end{itemize}

MotionTTT and stacked U-nets are  both trained on the Calgary Campinas Brain MRI Dataset~\cite{souzaOpenMultivendorMultifieldstrength2018} consisting of T1-weighted 3D motion-free brain scans recorded under a similar setting as our PMoC3D dataset. For MotionTTT and AltOpt, we estimate six motion parameters (three rotations, three translations) per acquired shot, and perform L1-minimization based on the estimated motion parameters for the final reconstruction. Shots with motion parameters that have a data consistency (DC) loss above a certain threshold are excluded from the measurements. See Appendix~\ref{appsec:hyper_details} for all training details and hyperparameters. 


\subsection{Assessing paired real-world evaluation}
\label{subsec:assess_real_paired}
Paired real-world evaluation is challenging because no truly motion-free ground truth exists. In PMoC3D, the reference volume is acquired in a separate scan and reconstructed via L1-minimization, and thus is not perfect ground-truth data. Moreover, it might be slightly misaligned to the motion-corrupted scan even if the motion-free scan had not motion corruption. 

\paragraph{Assessing evaluation reliability with a human judge.} 
Because the motion-free volume is not a perfect ground truth, we evaluate reliability by studying whether the ranking produced by different metrics measuring similarity of reconstructions and references based on the PMoC3D dataset agree with human perception.  

We performed exhaustive pairwise comparisons of all baseline reconstructions, following the protocol described in Section \ref{sec:pmoc_motion_severity}. The resulting preference matrix was fitted with a Bradley-Terry model (details in Appendix~\ref{appsec:L1_details}), which yields perceived motion artifact scores (PMAS).
We assessed the association between PMAS and each evaluation metric using Spearman's rank correlation coefficient. Correlations for PSNR and MoMRISim for medium and severe motion  
are displayed in Figure \ref{fig:quantitative_eval_correlation}; results for other  metrics as well as for mild motion appear in the Appendix~\ref{appsec:corr_analysis}.
Figure~\ref{fig:quantitative_eval_correlation} shows that PSNR and MoMRISim yield rankings consistant with the human judage (i.e., with the perceived motion artifact scores (PMAS)).  MoMRISim, in particular, attains the highest correlation which is 0.95. This strong association demonstrates that our evaluation is a faithful proxy for expert assessment under severe motion, confirming that paired real-world evaluation is reliable under medium and severe motion.

\begin{figure}
    \centering
    \includegraphics[width=\linewidth]{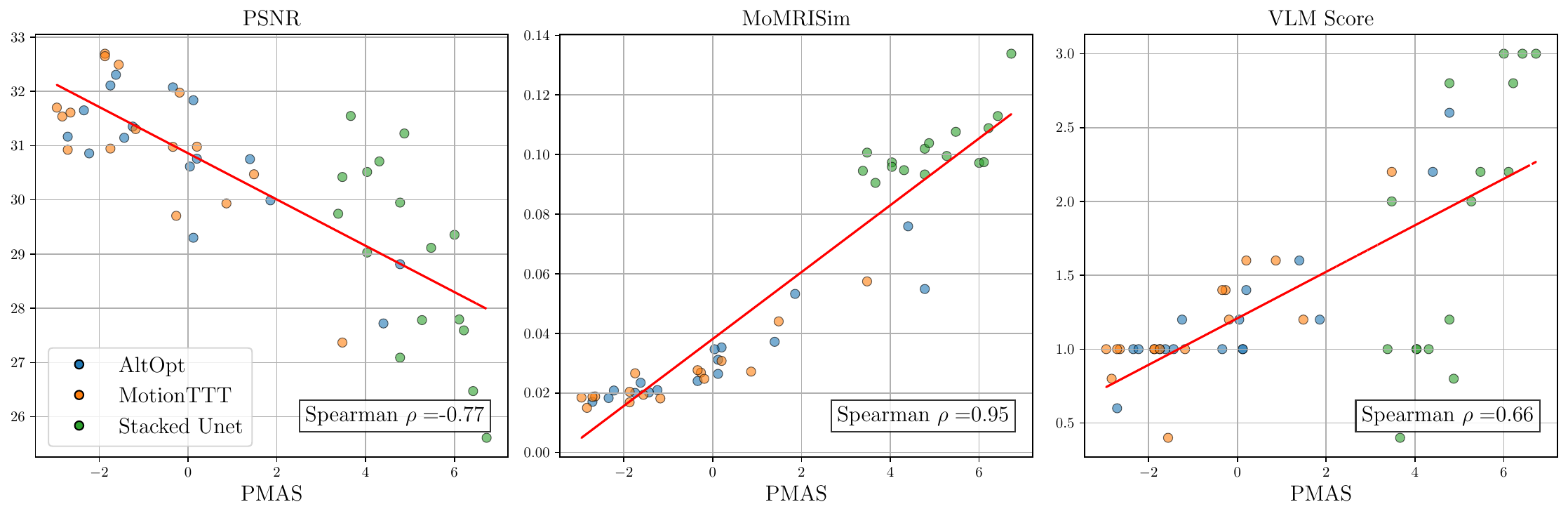}
    \caption{Correlation plot of PSNR$\uparrow$, MoMRISim$\downarrow$, and VLM score$\downarrow$ with the perceived motion artifact score.
    MoMRIScore shows a strong correlation, PSNR offers a moderate level, and VLM score reflects a low degree of alignment.
    } 
    \label{fig:quantitative_eval_correlation}
\end{figure}



\begin{figure}
    \centering
    \begin{tikzpicture}
        \node[anchor=south west] (image) at (0,0) {\includegraphics[width=0.68\linewidth]{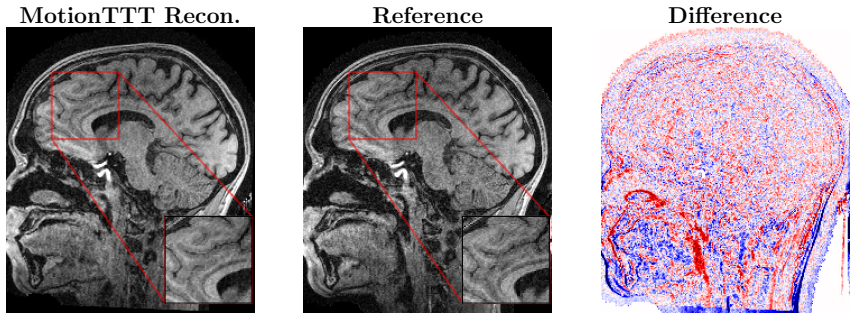}};
    \end{tikzpicture}
    \caption{Comparison of MotionTTT reconstruction, L1-based motion-free reference, and their difference image. The left panel shows the MotionTTT reconstruction of subject S3\_3, which is a mild motion scan. The center panel displays the L1 reconstruction of the corresponding motion-free reference. The right panel presents the difference image between the two reconstructions.
    }
    \label{fig:RefMotionMRI}
\end{figure}
\paragraph{Challenges in the paired real-world evaluation in particular for mild motion.} 
Figure~\ref{fig:RefMotionMRI} shows the motion-free reference image an the MotionTTT reconstruction of scan S3\_3, which was acquired under mild subject motion, along with the difference of the two images. For this example, the MotionTTT reconstruction of the mild motion scan is slightly better at parts than the reference image, which can be seen in the zoomed region: 
The MotionTTT reconstruction is free of ringing, whereas the L1 reconstructed motion-free reference contains faint ring artifacts. Under mild motion, the corrected image can therefore look better than the reference, which is problematic.

To summarize, although imperfect reference volumes may render paired real-world evaluation empirically unreliable in the mild-motion regime, real-world evaluation is dependable under moderate to severe motion: when combined with a robust reference-based metric such as MoMRISim, it yields performance assessments that closely align with human evaluation.

\subsection{Assessing simulated motion evaluation} 

Evaluation based on simulated motion is popular for its simplicity and since motion-corrupted real-world data is not required. However, such simulations may not reflect real-world motion artifacts sufficiently well. We find that evaluation based on simulated data overestimates the performance of reconstruction methods and can therefore be misleading.

\begin{figure}
    \centering
    \begin{tikzpicture}
        \node[anchor=south west] (image) at (0,0) {\includegraphics[width=0.78\linewidth]{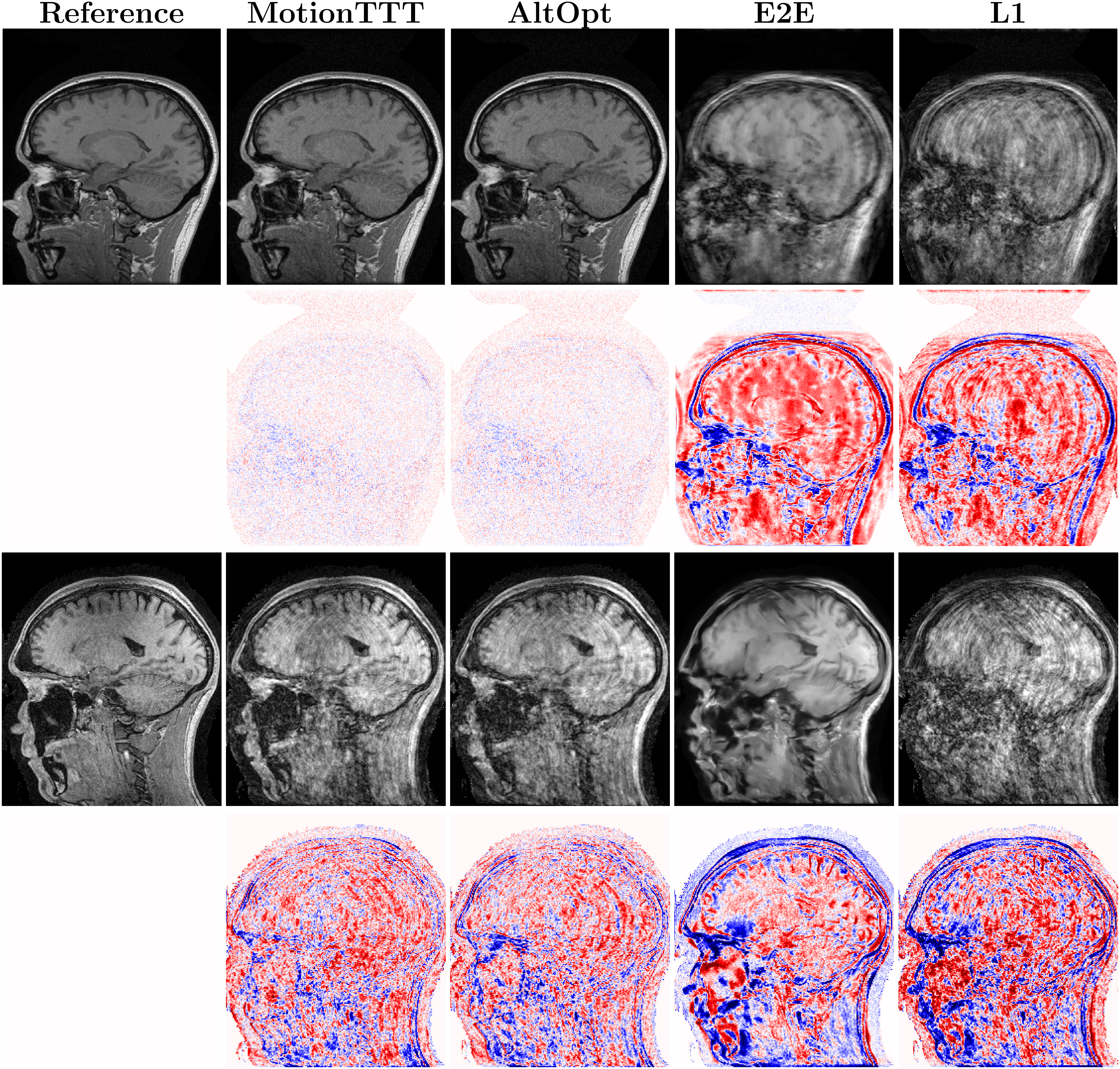}};
        
        \node[left, text width=15mm, align=left] at ([xshift=-1.0mm, yshift=4.5cm]image.west)  {\textbf{Simulated Recon.}:}; 
        \node[left, text width=30mm, align=left] at ([xshift=15mm, yshift=1.2cm]image.west)  {\textbf{Error w.r.t. reference}}; 
        \node[left, text width=15mm, align=left] at ([xshift=0mm, yshift=-1.8cm]image.west)    {\textbf{PMoC3D Recon.}:}; 
        \node[left, text width=30mm, align=left] at ([xshift=15mm, yshift=-4.5cm]image.west)  {\textbf{Error w.r.t. reference}}; 

    \end{tikzpicture}
    \caption{
        Baseline reconstructions of scans corrupted with simulated and PMoC3D artifacts and corresponding motion-free reference scans in the first column. Error maps are computed between reconstructions and the corresponding reference.  
        }
    \label{fig:CompareSimReal}
\end{figure}
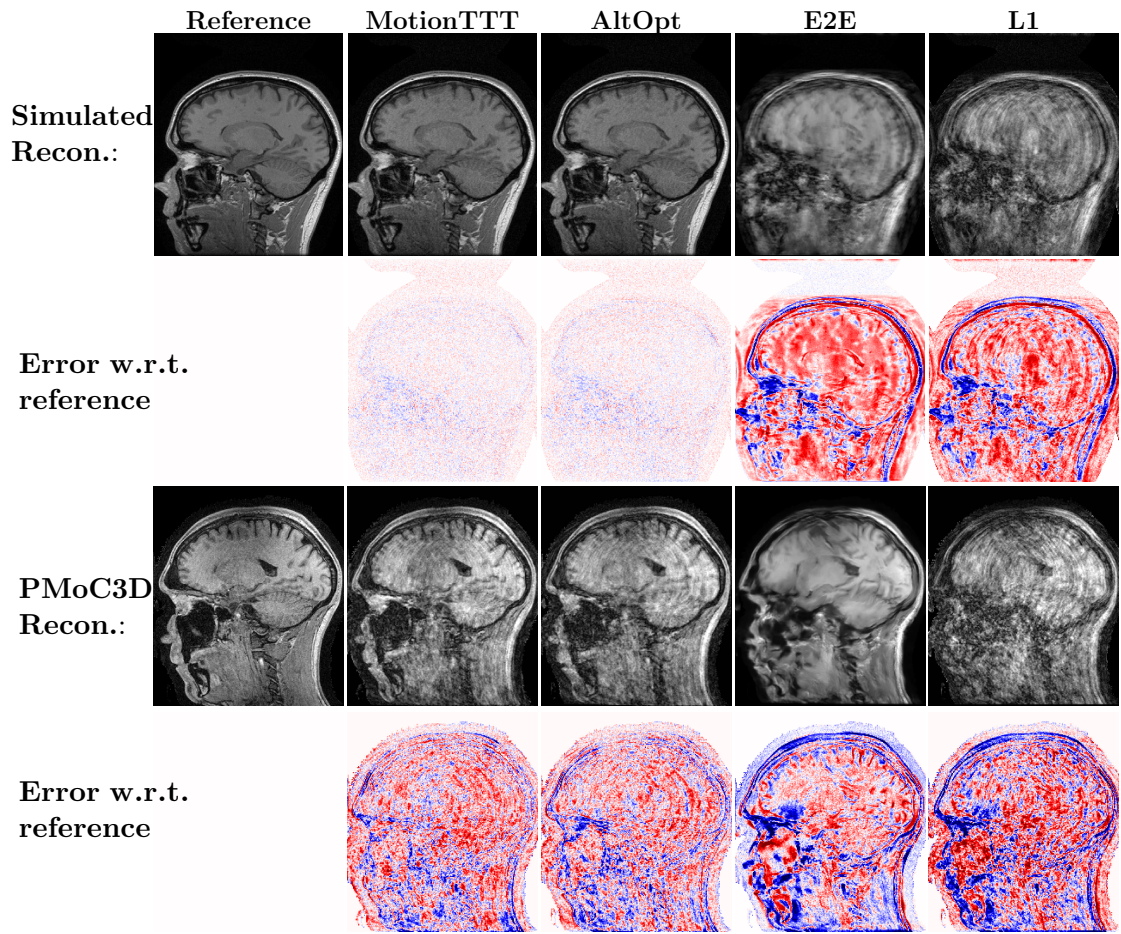

This is easy to see from Figure~\ref{fig:CompareSimReal}, which contains a severely corrupted real scan from the PMoC3D dataset as well as a simulated motion with a similar level of motion corruption (as seen in the L1 reconstructions in the figure). 
It can be seen that the reconstructions of all considered algorithms are significantly better for simulated data compared to the reconstructions based on the real data. This example is representative, and more examples of this are shown in Figure \ref{fig:compare_sim_real_large} in the appendix.

To underline this finding, we conducted a comparison experiment. 
We investigate whether a reconstruction method (we take MotionTTT) can better reconstruct simulated motion artifacts than real-world motion artifacts, even when they suffer from a similar level of motion corruption. We selected 8 volumes from the PMoC3D dataset with severe motion artifacts, and corrupted 8 motion-corrupted volumes with  simulated motion (as described in Section~\ref{subsec:eval_sim_desc}), with 
artifact severity closely resembling the real-world motion corrupted volumes.

We then created 64 comparison pairs by matching each real-world sample with every simulated sample. A human annotator conducted a blind evaluation of the MotionTTT reconstructions for each pair. The comparison focused on the brain regions, assessing which volume exhibited more motion artifacts or whether the two were similar in quality. The results showed that in 75\% of the comparisons, the real-world reconstructions exhibited clearly more residual artifacts than their simulated counterparts. The remaining 25\% were judged to have a similar level of motion artifacts.


Our comparison experiment confirms that, even at comparable levels of artifact severity, reconstructions from simulated motion consistently appear cleaner than those from real-world motion. 
This discrepancy indicates that evaluation approaches restricted to simulation therefore risk systematically overestimating algorithmic progress.

\subsection{Assessing reference-free evaluation} 

Reference-free evaluation can test for the presence of artifacts, but cannot measure  accuracy since no reference is available. We find that reference-free metrics systematically overestimate the performance of the deep learning based method.

To evaluate metric reliability, we examine the correlation between reference-free quality scores and PMAS. As illustrated in Figure~\ref{fig:quantitative_eval_correlation}, right pannel, the reference-free VLM metric exhibits a weak correlation with the human judage. A notable failure case involves certain stacked U-net reconstructions, where the VLM score assigns high quality despite PMAS indicating substantial motion artifacts. This discrepancy suggests that reference-free metrics do not reliably align with human perception of artifact severity. In Appendix~\ref{app:fail_ref_free}, we provide an example where a reference-free metric fails to reflect the quality of a reconstruction. Although the stacked U-net rated favorably by the metric, the corresponding difference image reveals substantial loss of anatomical detail. 


Reference-free evaluation is prone to bias, particularly for end-to-end deep learning models that suppress artifacts through oversmoothing.

\section{Conclusion}
\label{sec:conclusion}
Reliable evaluation of motion correction algorithms in 3D MRI is fundamentally difficult due to the absence of good ground truth in real-world data. This paper presents a comprehensive assesment of three evaluation approaches: real-world evaluation with a reference scan, based on simulated motion, and reference-free assessment.

To enable real-world evaluation, we introduced the PMoC3D dataset consisting of paired motion-free and motion-corrupted scans. 
We find that real-world evaluation is well correlated with human judgement of reconstruction and is thus relatively reliable. However,  for very mild motion, baseline reconstruction methods can produce better results than the motion-free reference, which potentially compromises the validity of reference-based evaluation in the mild-motion regime. 

Evaluation based on simulated motion can be misleading because simulated motion fails to capture the full complexity of real-world motion an tends to overestimate performance. However, evaluation based on simulated motion can still be useful for relative comparisons. 

Reference-free evaluation can be very biased towards certain reconstructions and is not reliable, as expected.

\subsubsection*{Reproducibility} 
The dataset can be found at \url{https://huggingface.co/datasets/mli-lab/PMoC3D} and the code for evaluation can be found at \url{https://github.com/MLI-lab/PMoC3D/tree/main}. The code to reproduce the baseline reconstruction is available at \url{https://github.com/MLI-lab/MRI_MotionTTT}. 

\subsubsection*{Acknowledgements}

This work was supported by the Deutsche Forschungsgemeinschaft (DFG,
German Research Foundation) - 456465471, 517586365 and the German Federal Ministry of Education
and Research, and the Bavarian State Ministry for Science and the Arts.

\printbibliography{}





\newpage
\appendix

\section{PMoC3D acquisition details}
\label{appsec:dataset_details}

In Table~\ref{tab:invivo_sequenceparameters} we provide a list of all relevant sequence parameters used for the acquisition of our PMoC3D dataset.
\begin{table}[ht]
  \caption{Sequence parameters of the PMoC3D dataset.}
  \label{tab:invivo_sequenceparameters}
  \centering
  \begin{tabular}{ll}
    \toprule
    Parameter     & Value \\
    \midrule
    Sequence & 3D T1-TFE \\
    Sampling  & Cartesian \\
    Flip angle (deg) & 8 \\
    TR (ms) & 6.7 \\
    TE (ms) & 3.0 (shortest) \\
    TFE prepulse / delay (ms) & non-selective invert / 1060 ms \\
    Min. TI delay (ms) & 707 \\
    TFE factor & 204 \\
    TFE shots & 52 \\
    TFE dur. shot / acq (ms) & 1742 / 1347 \\
    Shot interval (ms) & 3000 \\
    TFE prepulse delay (ms) & 1060 \\
    Under-sampling factor &  4.94 \\
    Half-scan factor Y / Z & 1 / 0.85 \\
    Number of auto-calibration lines & 37 \\
    Profile order & random \\
    Field of view (FH x AP x RL, mm)  & 256 x 221 x 170 \\
    Acquisition matrix  & 256 x 221 \\
    Fold-over direction & AP \\
    Fold-over suppression & no \\
    Fat shift direction & F \\
    Water-fat shift (pixels) & 1.6 \\
    Saturation slabs & no \\
    \bottomrule
  \end{tabular}
\end{table}

Figure~\ref{fig:data_display} illustrates an example from the sampled dataset, including the image volume, its corresponding k-space representation, and the undersampling mask pattern applied along 2 phase encoding directions.

{

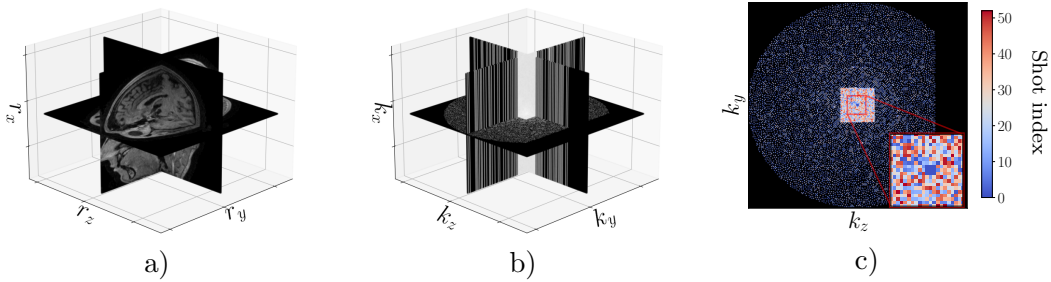
\begin{figure}[h]
    \centering
    \begin{tikzpicture}
    \def\imgwidth{4cm}
    \def\maskwidth{4cm}
    \def\labelfont{\normalsize}
    \node[] (3dimage) {\includegraphics[width=\imgwidth]{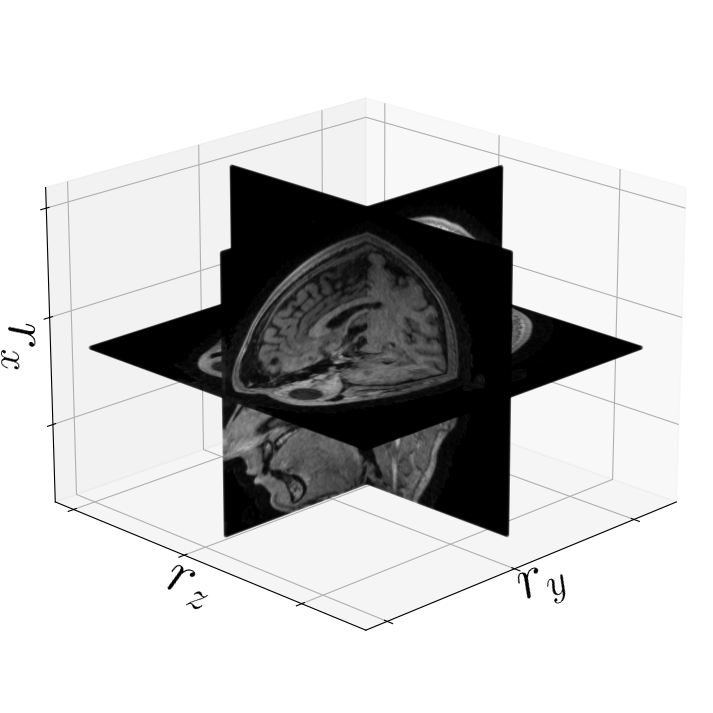}};
    \node[shift=(3dimage.south), anchor=north, yshift=5mm, xshift=0] (3dimagelab) {\labelfont a)};
    \node[shift=(3dimage.east), anchor=west,yshift=0,xshift=6mm] (3dksp) {\includegraphics[width=\imgwidth]{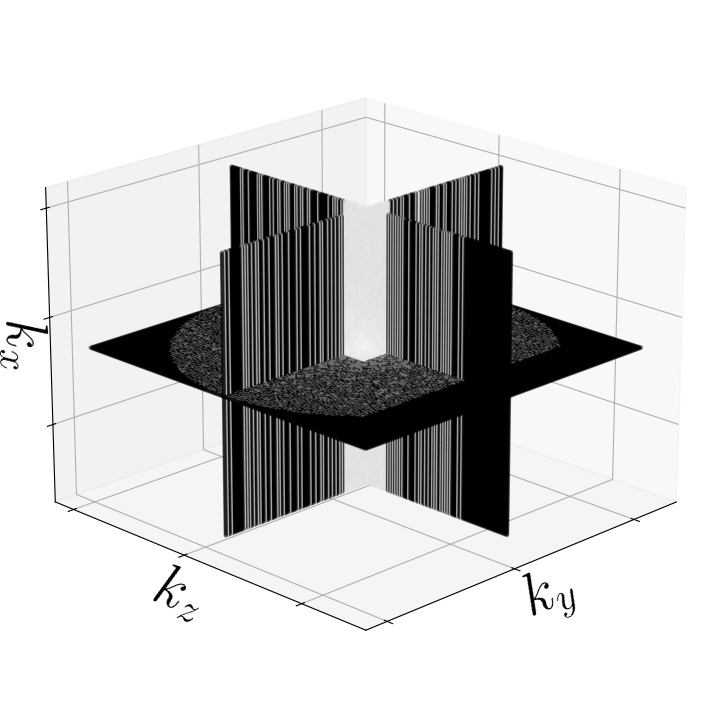}};
    \node[shift=(3dksp.south), anchor=north, yshift=5mm, xshift=0] (3dksplab) {\labelfont b) };
    
    \node[shift=(3dksp.east), anchor=west,yshift=0,xshift=3mm] (masksim) 
    {\includegraphics[width=\maskwidth]{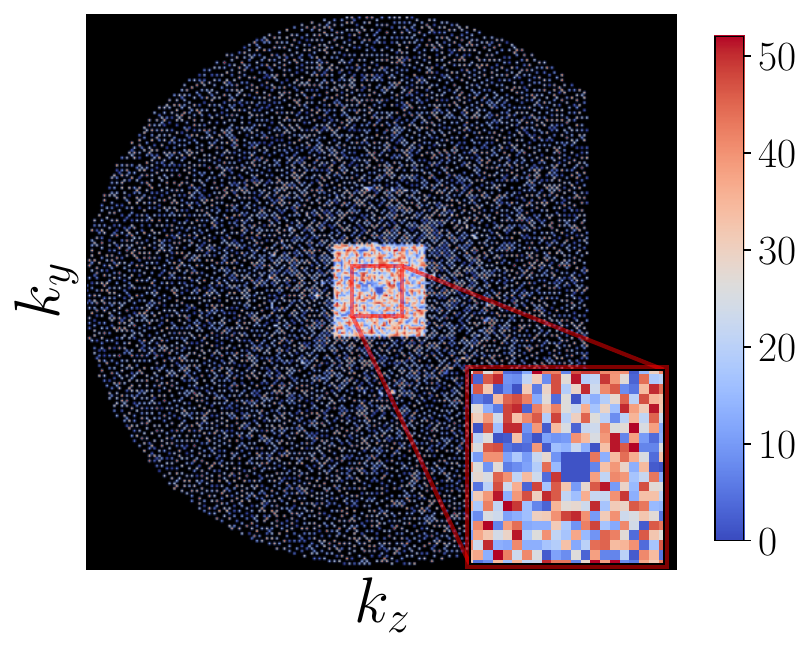}};
    \node[shift=(masksim.south), anchor=north, yshift=1.5mm, xshift=0] (masksimlab) {\labelfont c)};
    \node[rotate=-90, anchor=north,] at ([xshift=0.4cm,yshift=0.2cm]masksim.east) {\scriptsize Shot index};
    \end{tikzpicture}
    
    \caption{
    Panel a): schematic visualization of the magnitude of a 3D volume; Panel b): the corresponding 3D k-space data. Panel c): the undersampling masks with the color coding illustrating an example of a random sampling trajectory indicating which lines along the readout dimension $\dimz$ are sampled within the same out of 52 shots. 
    }
    \label{fig:data_display}
\end{figure}
}

\section{Perceived motion artifact score details}
\label{sec:hps_cal}
In order to evaluate the severity of motion artifacts in the L1 reconstructions, we first shuffle the reconstructions and conceal their labels. Then, the first two authors of this paper perform pairwise comparisons between the reconstructions of the 24 motion-corrupted scans. If both evaluators agree that reconstruction A has more severe artifacts than B then we assign rate $p(A>B)=1$. If one evaluator judges A to be better and the other finds a similar level, then we assign $p(A>B) = 0.75$. If both evaluators find a similar severity level, or one finds one better and the other the other, we set $p(A>B) = 0.5$. 

For the PMAS score described in Section~\ref{subsec:assess_real_paired}, which is used to compare reconstruction quality across different baselines and scans, annotations were performed by a single evaluator. The evaluation also follows a pairwise comparison protocol. For each pair of reconstructions, if the annotator judges that reconstruction A exhibits more severe artifacts than B, we assign a rate of $p(A>B)=1$. If the two reconstructions are considered to have similar artifact severity, the win rate is set to $p(A>B) = 0.5$.

Based on those pairwise comparisons, we fit a Bradley--Terry model~\cite{bradley1952rank}, which assigns a latent parameter to each reconstruction. The difference in these parameters indicates which volume is considered to have more severe motion artifacts. 
We estimated these parameters by maximizing the likelihood function 
\begin{equation}
\text{PMAS} = \arg\max_{\mathbf{\beta}}\sum_{i\neq j}p(i>j)\log\left(\frac{\exp(\beta_i)}{\exp(\beta_i)+\exp(\beta_j)}\right),
\end{equation}
using gradient descent with the Adam optimizer in PyTorch. In this formulation, each $\beta_i$ quantitatively represents the severity of motion artifacts for the corresponding volume; higher values indicate more severe artifacts, and these latent parameters serve as our perceived motion artifact score.

Table~\ref{tab:HPS_scores} presents each motion-corrupted scan's perceived motion artifact score in the PMoC3D dataset. Based on the perceived motion artifact score, we categorize the scans into three motion severity levels as follows:

\begin{itemize}
    \item \textbf{Mild Motion}: S1\_2, S7\_3, S7\_1, S3\_3, S1\_3, S4\_2, S5\_2, S2\_3
    \item \textbf{Severe Motion}:  S3\_2, S4\_1, S7\_2, S5\_3, S6\_2, S2\_2 ,S8\_2, S6\_1, S4\_3, S3\_1, S5\_1, S1\_1, S8\_1, S6\_3, S8\_3, S2\_1
\end{itemize}

This classification facilitates a structured analysis of the reconstruction methods' performance across varying degrees of motion artifacts.

\begin{table}[ht]
    \centering
    \caption{Perceived motion artifact score (PMAS) for each motion-corrupted scan in the PMoC3D dataset. Higher scores indicate more severe motion artifacts.}
    \label{tab:HPS_scores}
    \begin{tabular}{ll|ll|ll|ll}
        \toprule
        \textbf{Scan ID} & \textbf{PMAS} & \textbf{Scan ID} & \textbf{PMAS} & \textbf{Scan ID} & \textbf{PMAS} & \textbf{Scan ID} & \textbf{PMAS} \\
        \midrule
        S2\_1 &  2.417 & S4\_3 &  1.552 & S4\_1 &  0.808 & S7\_1 & -0.156 \\
        S8\_3 &  2.230 & S6\_1 &  1.405 & S3\_2 &  0.771 & S7\_3 & -0.231  \\
        S6\_3 &  2.197 & S8\_2 &  1.400 & S2\_3 &  0.485 & S1\_2 & -0.443   \\
        S8\_1 &  2.189 & S2\_2 &  1.102 & S5\_2 &  0.356 &  &  \\
        S1\_1 &  1.828 & S6\_2 &  1.041 & S4\_2 &  0.346 &  &   \\
        S5\_1 &  1.782 & S5\_3 &  0.867 & S1\_3 & -0.008 &  &    \\
        S3\_1 &  1.673 & S7\_2 &  0.813 & S3\_3 & -0.057 & &      \\
        \bottomrule
    \end{tabular}
\end{table}
\section{Hyperparameter configurations and implementation details of baselines}
\label{appsec:hyper_details}

In this Section, we provide further details regarding implementation and hyperparameter configurations used for reconstructing the PMoC3D dataset with classical L1-minimization~\cite{lustigCompressedSensingMRI2008}, alternating optimization similar to \cite{corderoSensitivityEncodingAligned2016}, MotionTTT~\cite{klugMotionTTT2DTestTimeTraining2024} and E2E Stacked U-net~\cite{almasniStackedUNetsSelfassisted2022} in Sections~\ref{sec:pmoc_motion_severity} and~\ref{subsec:baselines}.

\subsection{L1-minimization}
\label{appsec:L1_details}

We perform L1-minimization with the mean-squared-error loss function and wavelet regularization. We use the Haar wavelet implementation of order one from the \texttt{PyWavelets} package~\cite{leePyWaveletsPythonPackage2019} with PyTorch support through the \texttt{PyTorch Wavelet Toolbox}~\cite{wolterPtwtPyTorchWavelet2024}. 
We run 40 steps with stochastic gradient descent, a learning rate of $10^{8}$ and regularization weight $\lambda=3 \times 10^{-8}$.

\subsection{Alternating optimization}
\label{appsec:altopt_details}

To perform alternating optimization as described in Section~\ref{subsec:baselines} we run SGD with a learning rate of $10^{8}$ and regularization weight $\lambda=3^{-8}$ during the reconstruction steps and a learning rate of $5\times 10^{-2}$ during the motion estimation step. In both steps the loss is the MSE between predicted and given measurement. 
The optimization process is capped at 500 iterations, but it terminates early if the difference between the losses of the first and second reconstruction step is less than 0.02. 

After alternating optimization we perform L1-minimization from scratch based on the estimated motion parameters as describe in Section~\ref{appsec:L1_details}. 

\subsection{MotionTTT} 
\label{appsec:motionTTT_details}

To conduct test-time-training motion estimation with MotionTTT~\cite{klugMotionTTT2DTestTimeTraining2024} we use the model provided by the authors, which is pre-trained on the Calgary Campinas Brain MRI Dataset~\cite{souzaOpenMultivendorMultifieldstrength2018} for the task of 2D motion-free reconstruction from undersampled MRI. 

As outlined in~\cite{klugMotionTTT2DTestTimeTraining2024} the iterative motion estimation can be conducted in three phases, where during phase 2, motion states pertaining to shots that exhibit a large data consistency (DC) loss, can be split into several distinct motion states to estimate a more fine-grained motion trajectory during phase 2 and 3.
This can improve the reconstruction quality compared to terminating MotionTTT after phase 1 as potentially less measurements have to be discarded during the DC loss thresholding before the final reconstruction. 

For the PMoC3D dataset we observed no significant difference between the reconstruction quality of splitting corrupted shots during phase 2 or terminating the optimization after phase 1 and directly thresholding the corrupted shots from the reconstruction. 
The number of read-out lines that are saved from being thresholded during phase 2 lies in the range from 1-5\% of the total number of lines, which appears to be too little to make a visual difference in the reconstruction. 

Hence, for the results discussed here we reduce the computational costs of MotionTTT by running only phase 1, where one motion state (3 rotation and 3 translation parameters) is estimated per acquired shot. Specifically, we run 80 steps with an initial learning rate of 1.0 reduced by a factor of 2 at steps 50, 60 and 70. All other parameters are set as in~\cite{klugMotionTTT2DTestTimeTraining2024}. 

For the final reconstruction we run L1-minimization as described in Section~\ref{appsec:L1_details} based on the estimated motion parameters, where shots with a DC loss larger than a threshold of 0.70 are excluded from the reconstruction. 

\subsection{E2E Stacked U-net}
\label{appsec:E2E_details}

For the E2E Stacked U-net baseline results we adopt the network design from~\cite{almasniStackedUNetsSelfassisted2022}, where we set the number of channels in first layer of both U-nets to 64 resulting in a total of 15.9M network parameters. We used instance norm instead of batch norm in our network as we found it to give more stable results. 

We train the model on the slices of 40 volumes from the Calgary Campinas Brain MRI Dataset~\cite{souzaOpenMultivendorMultifieldstrength2018}. 
In every training step one 3D volume is loaded from which the fully sampled target volume is computed as well as motion-free and motion-corrupted undersampled input volumes. 
Then, 20/10 slices are selected randomly from the motion-free/motion-corrupted volumes in each plane $\numPd\times\numQd$, $\numPd\times\numFd$ and $\numQd\times\numFd$ together with the corresponding target slices resulting in a total of 90 input-target pairs per training step. Thus, with a batch size of 10 the network parameters are updated 9 times per training step and $40*9=360$ times per epoch. 
We train the model for 200 epochs with the SSIM loss and the Adam optimizer with a learning rate of $6 \times 10^{-4}$ which is decayed twice by a factor of 10 at epochs 130 and 170.

We use twice as many slices from the motion-free input to ensure that the network can achieve high quality reconstructions in the absence of motion. 
We generate the motion-corrupted volumes based on the inter-shot motion simulation model from \cite{klugMotionTTT2DTestTimeTraining2024}, where we focus on very mild motion with either one or two motion events. 
During a motion event 1-6 randomly selected motion parameters are set to a value drawn uniformly from either $[-1,1]$ degrees/mm or $[-2,2]$ degrees/mm simulating subject movement in between the recording of two shots. 
We focus the model training on mild motion as for more severe artifacts image details are occluded and thus irreversible lost for reconstruction with an end-to-end approach. 
Nevertheless, we note that the motion correction capability of an end-to-end model is specific to the type of motion simulated during training and hence a more sophisticated motion simulation could benefit the model's performance in the regime of mild motion. 


\section{Implementation of MoMRISim}
\label{app:MoMRISim}
We propose a motion MRI similarity(MoMRISim) as a learned perceptual metric to quantify motion-artifact severity in 3D MRI. The basic idea of the MoMRISim is described in Section~\ref{subsec:eval_metric}. We are going to describe the implementation details on this section. 

We train on 40 brain volumes from the Calgary Campinas (CC-359) dataset. For each training volume we apply rigid-body motion corruption to the k-space under 7 randomly sampled severity levels, where severity definitions-ranging from mild to severe in terms of number of motion events and rotation/translation perturbation magnitudes, with definitions following \cite{klugMotionTTT2DTestTimeTraining2024}. 
And reconstruct each corrupted k-space without motion correction by both an L1-wavelet reconstruction and a 2D U-Net. This ensures that MoMRISim observes artifact patterns from both classical compressed sensing and deep learning pipelines.

\begin{figure}
    \centering
    \begin{tikzpicture}
        \node[anchor=south west] (image) at (0,0) {\includegraphics[width=0.8\linewidth]{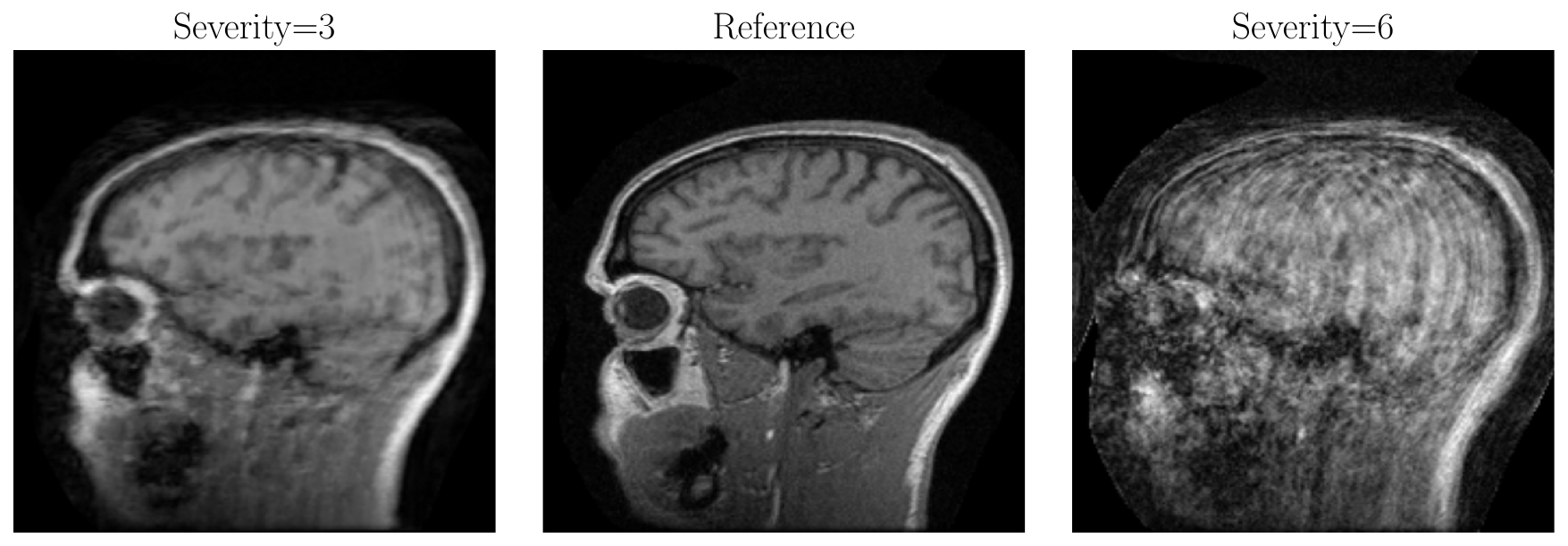}};

    \end{tikzpicture}
    \caption{
Example triplet from the MoMRISim training dataset. The left image is corrupted by motion (severity level 3) and reconstructed using a 2D U-Net. The right image is corrupted by more severe motion (level 6) and reconstructed with an L1 reconstruction without motion correction. The center image is a motion-free reconstruction using an L1 reconstruction.
        }
    \label{fig:triplet_example}
\end{figure}

Training samples are triplets $\bigl\langle \text{Ref},\,C_{1},\,C_{2}\bigr\rangle$,
where Ref denotes the L1-wavelet reconstruction of the original motion-free volume, and \(C_{1}\), \(C_{2}\) are reconstructions of the same slice under two distinct motion severities. For example, if the severity of \(C_{1}\) lower than that of \(C_{2}\), then Ref should be closer to \(C_{1}\). An example of the triplet is shown in Figure~\ref{fig:triplet_example}.

In each epoch, we construct triplet by enumerating all pairs of reconstructions at two distinct severity levels among the seven corrupted volumes.
And then randomly select one of the three anatomical planes (axial, coronal, sagittal), sample ten slices per pair, normalize each slice by its 99.9th-percentile intensity, and discard any background-only slices. This yields approximately 7 000 triplets per epoch.

We adapted the same training way as the DreamSim\cite{fu2023dreamsim} to fine-tune the DINO-vitb16\cite{DBLP:journals/corr/abs-2104-14294} visual encoder augmented with LoRA adapters by minimizing a hinge-ranking loss. 
The model was optimized with AdamW (learning rate $3.0\times10^{-5}$), LoRA rank 4, and a batch size of 8, over 40 epochs. Training was conducted on NVIDIA RTX A6000 GPUs and completed in approximately 90 minutes with 4 workers. The final model achieved a triplet-ranking accuracy of 0.933 on the training set.

\section{Implementation of  VLM score}
\label{sec:gpt_evaluation}
\begin{figure}
    \centering
    \begin{tikzpicture}
        \node[anchor=south west] (image) at (0,0) {\includegraphics[width=0.45\linewidth]{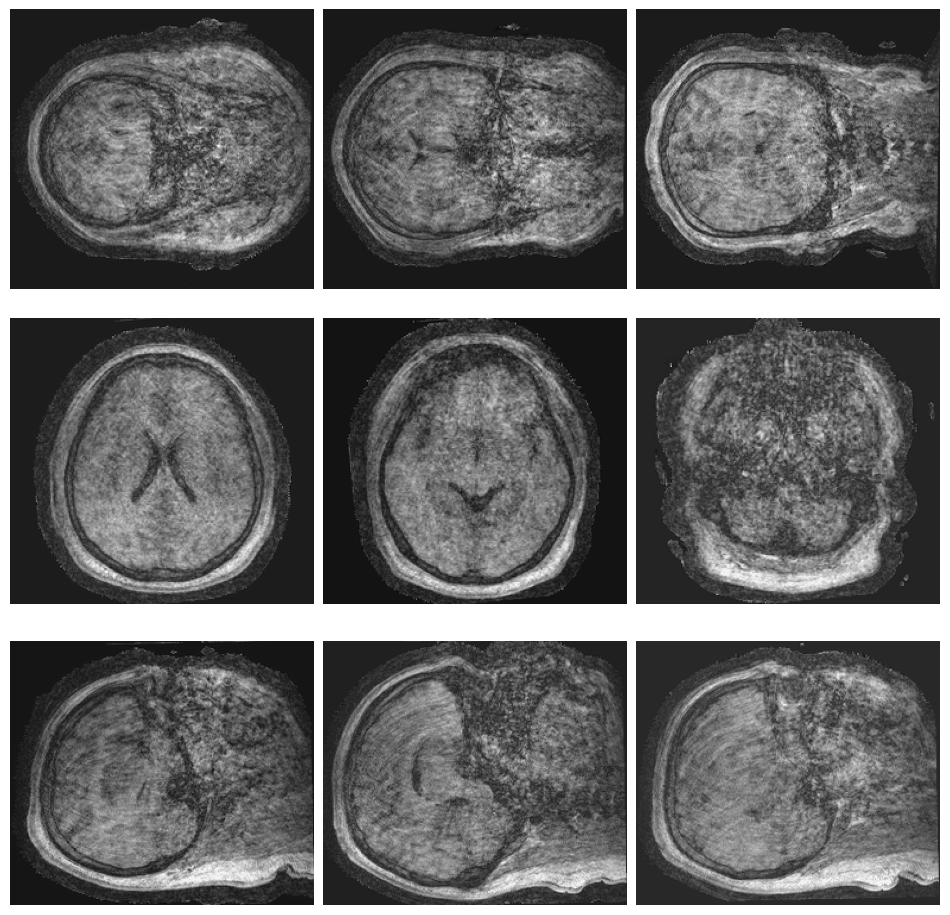}};
    \end{tikzpicture}
    \caption{
    An example input for VLM evaluation, where three slices from sagittal, coronal, and axial orientations are arranged in a 3x3 grid for assessment.
    }
    \label{fig:example_gpt_eval}
\end{figure}

Vision-language models (VLMs) have found extensive applications across various domains. In this study, we employ GPT-4o \cite{openai_gpt4o,openai2024gpt4ocard} to evaluate reconstruction quality. First, we input only the reconstructed images into GPT-4o, thus using it as a reference-free metric. To enable GPT-4o to have a better view of the images, we arrange three slice images from three different views into a 3x3 grid as the input for GPT-4o. Figure~\ref{fig:example_gpt_eval} provides an example of this setup.

To ensure consistent and reliable evaluations, we set the temperature parameter of GPT-4o to 0.5, which balances diversity and coherence in the generated outputs. For each evaluation instance, we provide a prompt to GPT-4o, generating five independent responses. Each response is categorized by GPT-4o into one of four predefined levels: No Motion, Mild, Moderate, or Severe, denoted as scores 0, 1, 2, and 3 respectively. The final evaluation score for each instance is calculated as the average of these five categorizations. The prompt we used is as follows:
\begin{tcolorbox}[colback=white, colframe=black, arc=2mm]
\lstset{
    basicstyle=\footnotesize\ttfamily, 
    breaklines=true,
    frame=none, 
}
\begin{lstlisting}
**Task:**  
Evaluate the severity of motion artifacts in the provided MRI image using a structured and systematic analysis.
---
### **Evaluation Criteria for MRI Image**
- **No Motion Artifact:** No visible motion artifacts; excellent diagnostic quality, and minor reconstruction noise is acceptable.
- **Mild:** The majority of brain details are clearly visible, with only minor artifacts that do not obscure diagnostic structures; minimal diagnostic impact, and minor reconstruction noise is acceptable.
- **Moderate:** Noticeable artifacts that partially obscure critical diagnostic regions; artifacts significantly impact diagnostic interpretation.  
- **Severe:** Brain structures are predominantly obscured by artifacts, with only the general shape discernible; diagnosis is extremely challenging or impossible.
### **Output Template**
**Analyze Brain Structure Visibility**  
   - Does the image look very smooth, potentially losing significant detail? *(Important for scoring!)*  
   - Are all major brain details visible (gyri, sulci, ventricles)?  
   - Do motion artifacts blur or distort critical brain details?  
   - Are there regions where brain details are completely lost?
**Assess Artifact Types and Locations**  
   - Check for ringing effects (where, how severe).  
   - Identify other motion artifacts (streaking, ghosting) and note their severity.
**Oversmooth Assessment**  
    - Does the image look very smooth (like a very high-quality image)?  
    - Are there areas with smooth distortions?  
    - If yes, do you think the image has an oversmoothing problem?
- The primary MRI image shows **[overall assessment]** motion artifacts. The final precise motion artifact level is: [No Motion/Mild/Moderate/Severe]
If the severity level is No Motion/Mild: Re-examine the image. Are all details truly clear? If any structures appear compromised, consider increasing the severity level.
---
### **Conclusion**
- After rethinking, the primary MRI image shows **[overall assessment]** motion artifacts, and the details are **. Given these factors, the final precise motion artifact level is:
Severity Level: [No Motion/Mild/Moderate/Severe]
\end{lstlisting}
\end{tcolorbox}

\section{Additional experiment results}
\label{app:add_res}

\subsection{Correlation analysis between PMAS and metrics on PMoC3D reconstructions}
\label{appsec:corr_analysis}
This section presents the correlation results between the Perceived Motion Artifact Score (PMAS) and various image quality metrics computed on the PMoC3D reconstructions. The correlations are the figures for all mild, moderate, and severe situation.

\begin{figure}
    \centering
    \begin{tikzpicture}
        \node[anchor=south west] (image) at (0,0) {\includegraphics[width=\linewidth]{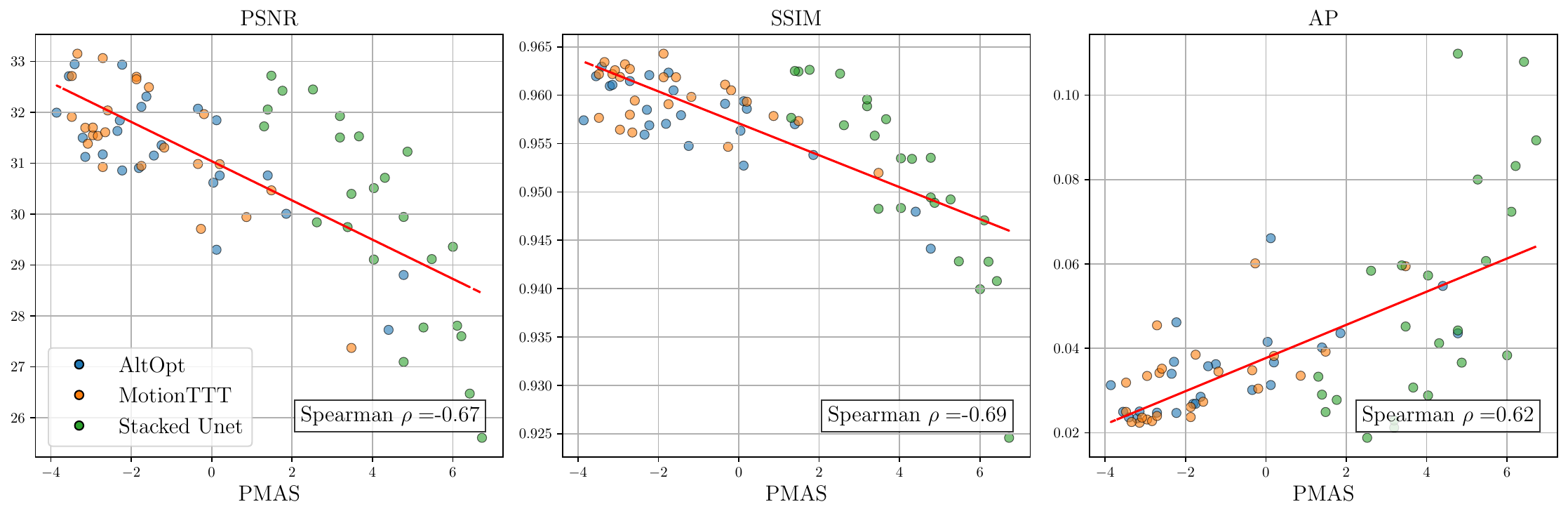}};
    \end{tikzpicture}
    \caption{Correlation plot of PSNR$\uparrow$, SSIM$\uparrow$, and AP$\downarrow$ with the perceived motion artifact score.
    All of them show moderate correlation with human judgment.
    }
    \label{fig:corr_classic}
\end{figure}

Figure~\ref{fig:corr_classic} shows the correlation between PMAS and traditional pixel-wise metrics. While these metrics generally reflect the expected trend of increasing degradation with higher motion artifacts, their correlation with PMAS is only moderate. This suggests that pixel-wise metrics have limited sensitivity to perceptual quality differences caused by motion corruption and may not be fully reliable for evaluating motion-degraded 3D MRI.

\begin{figure}
    \centering
    \begin{tikzpicture}
        \node[anchor=south west] (image) at (0,0) {\includegraphics[width=\linewidth]{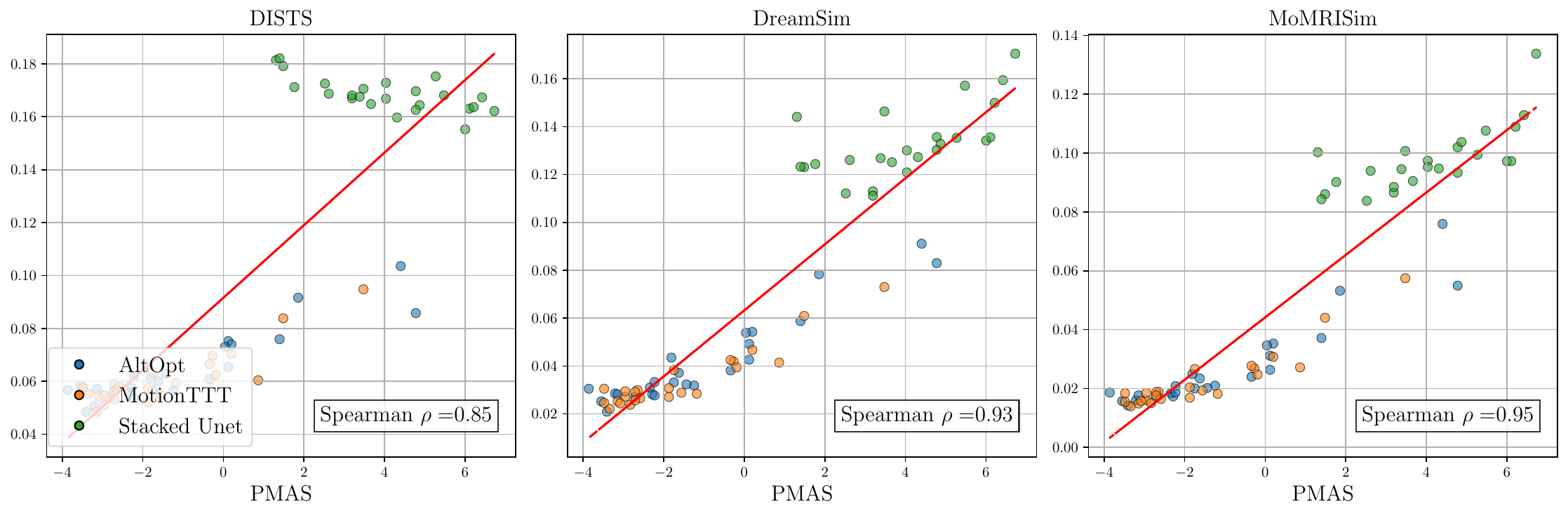}};
    \end{tikzpicture}
    \caption{Correlation plot of DISTS$\downarrow$, DreamSim$\downarrow$, and MoMRISim$\downarrow$ with the perceived motion artifact score.
    All of them show high correlation with human judgment, while the MoMRISim shows the highest correlation.
    }
    \label{fig:corr_feature}
\end{figure}

Figure~\ref{fig:corr_feature} presents the correlation between PMAS and feature-based metrics. Overall, these metrics show a strong correlation with PMAS. However, DISTS demonstrates a notable failure mode when evaluating reconstructions from the stacked U-net, consistently assigning them abnormally low scores. In contrast, both DreamSim and MoMRISim exhibit more stable behavior and higher alignment with PMAS. Among them, MoMRISim achieves the highest correlation with PMAS, indicating its robustness in capturing motion artifacts.

\begin{figure}
    \centering
    \begin{tikzpicture}
        \node[anchor=south west] (image) at (0,0) {\includegraphics[width=\linewidth]{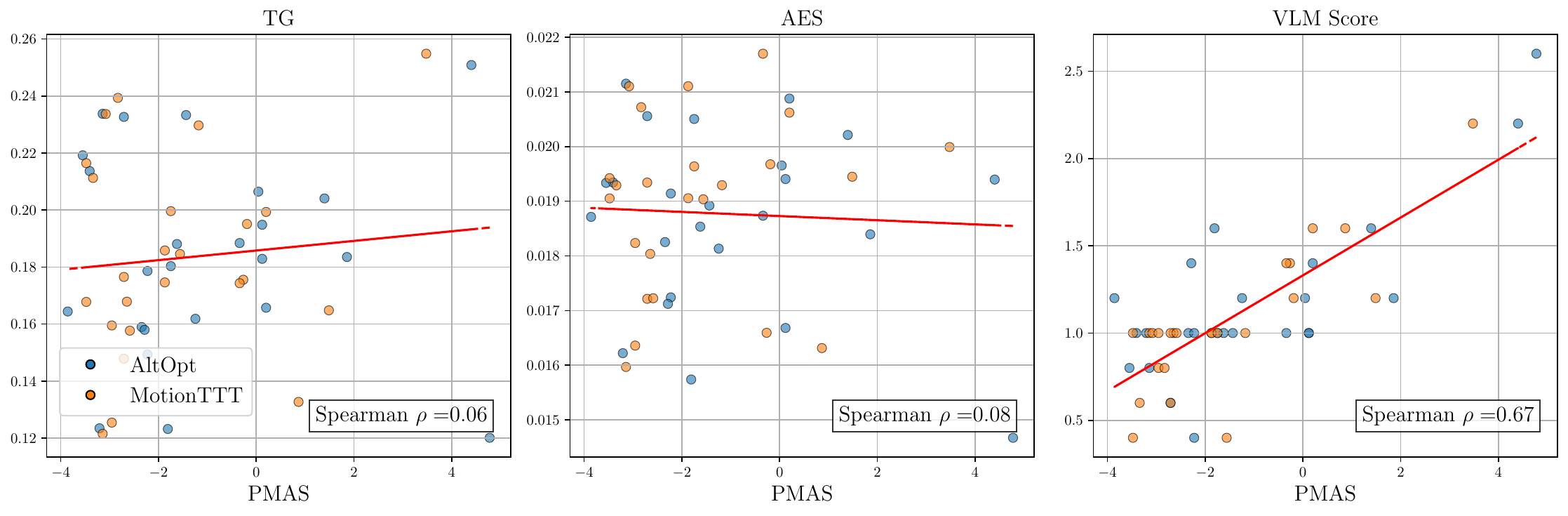}};
    \end{tikzpicture}
    \caption{Correlation plot of TG$\uparrow$, AES$\uparrow$, and VLM Score$\downarrow$ with the perceived motion artifact score.
    All of them show poor correlation with human judgment.
    }
    \label{fig:corr_ref_free}
\end{figure}

Figure~\ref{fig:corr_ref_free} presents the correlation between PMAS and reference-free metrics, which overall show poor alignment with human judgment. This analysis focuses specifically on reconstructions from AltOpt and MotionTTT, both of which use L1 minimization, to ensure a consistent reconstruction baseline. Among the evaluated metrics, the VLM score demonstrates significantly better alignment with PMAS than TG and AES. These results suggest that while reference-free metrics are generally less reliable for assessing motion artifacts, the VLM score may offer a practical alternative when references are unavailable.

\subsection{Comparison between simulated and real-world evaluation under severe motion}
\label{appsec:compar_sim_real_severe}
\begin{figure}
    \centering
    \begin{tikzpicture}
        \node[anchor=south west] (image) at (0,0) {\includegraphics[width=0.78\linewidth]{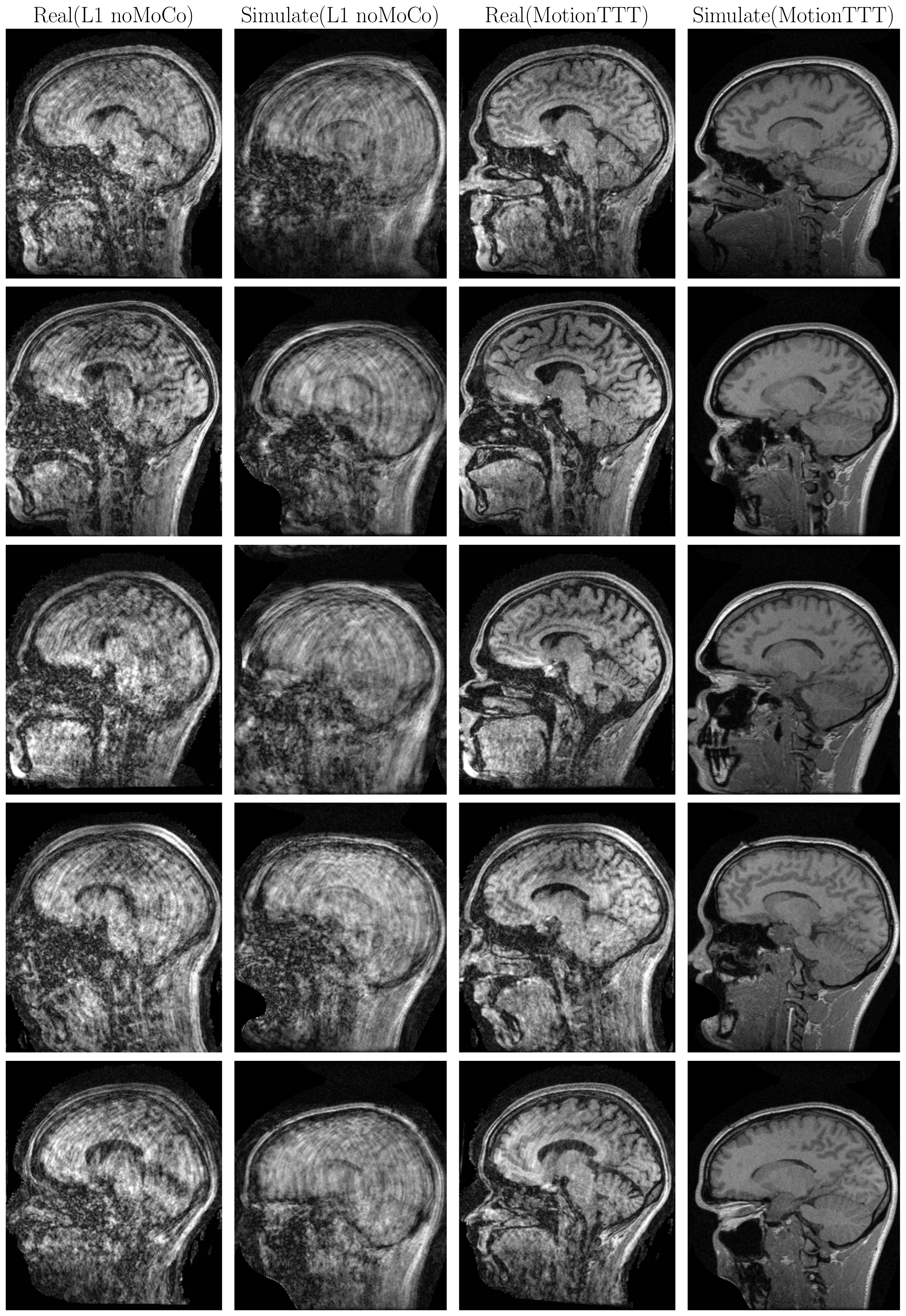}};
    \end{tikzpicture}
    \caption{
Reconstructions from the PMoC3D (real) and simulated datasets under two reconstruction methods. The first two columns show L1 reconstructions without motion correction, where simulated volumes exhibit motion artifacts of similar or greater severity compared to real-world scans. The last two columns display MotionTTT reconstructions: while real scans retain visible motion artifacts, simulated volumes are consistently corrected with minor residual artifacts.
        }
    \label{fig:compare_sim_real_large}
\end{figure}

Figure~\ref{fig:compare_sim_real_large} shows five sets of reconstructions with and without motion correction under severe motion artifacts, comparing PMoC3D (real) and simulated motion cases. The first two columns show L1 reconstructions without motion correction, which reflect the raw severity of motion artifacts. Both real and simulated scans display strong artifacts. Notably, the real-world L1 reconstructions still preserve some anatomical details, while the simulated counterparts often obscure brain structures entirely-indicating that the simulated artifact severity is comparable to or even greater than that of real-world scans.

The last two columns present the corresponding MotionTTT reconstructions. In all cases, real-world data retains noticeable ringing artifacts, with the fifth row showing particularly obvious artifacts. In contrast, the simulated MotionTTT reconstructions appear consistently clean, with motion artifacts largely eliminated. 

Given that the simulated artifacts are at least as severe as those in real-world scans, the significantly better reconstruction quality further confirms that simulation-based evaluation can lead to a systematic overestimation of reconstruction performance in practical settings.
 
\subsection{Failure example of reference-free evaluation}
\label{app:fail_ref_free}
\begin{figure}
    \centering
    \begin{tikzpicture}
        \node[anchor=south west] (image) at (0,0) {\includegraphics[width=0.85\linewidth]{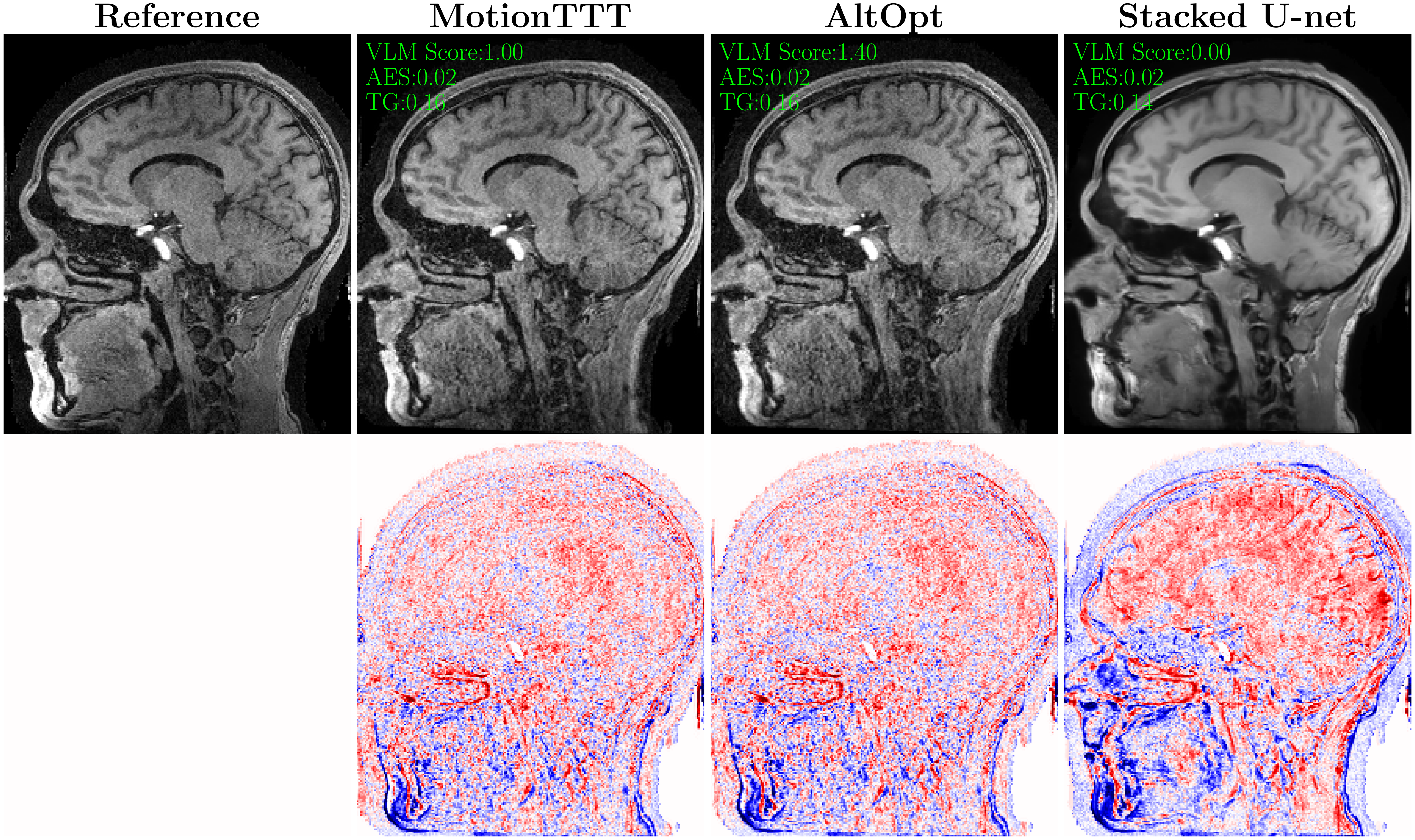}};
        
        \node[left, text width=15mm, align=left] at ([xshift=-0.1cm, yshift=1.7cm]image.west)  {\textbf{Simulated Recon.}:}; 
        \node[left, text width=30mm, align=left] at ([xshift=15mm, yshift=-2.2cm]image.west)  {\textbf{Error w.r.t. reference}}; 

    \end{tikzpicture}
    \caption{
        Baseline reconstructions of scan S4\_2, with the difference images and the calculated reference-free scores.
        }
    \label{fig:MRI_failure_ref_free}
\end{figure}

In this section, we present an additional failure case of reference-free evaluation. Figure~\ref{fig:MRI_failure_ref_free} shows reconstructions from different baselines along with their corresponding error maps.

As illustrated, the stacked U-Net reconstruction exhibits substantial loss of anatomical detail, as clearly visible in the error image. This result is qualitatively worse than those produced by MotionTTT and AltOpt. However, due to its visually smooth appearance, the stacked U-Net receives comparable scores from TG and AES, despite its degraded quality.

Even the VLM score while generally better aligned with human judgment fails in this case, assigning a near-perfect score to the stacked U-Net reconstruction. This example underscores a key limitation of reference-free metrics: they can be misled by superficially clean outputs that actually lack critical structural fidelity.

\end{document}